\documentstyle[numreferences,epsf]{kluwer}

\runningtitle{Nonlinear Gravitational Clustering in  Expanding Universe}

\begin{opening}

\title{Nonlinear Gravitational Clustering in  Expanding Universe}
\author{T. \surname{Padmanabhan }}
\institute{Inter-University Centre for Astronomy and Astrophysics\\
Post Bag 4, Ganeshkhind\\
Pune - 411 007}

\end{opening}

\begin{ao}
T. Padmanabhan,
IUCAA,
Post Bag 4, Ganeshkhind, Pune - 411 007.
Fax: (212)350760, Email: paddy@iucaa.ernet.in
\end{ao}

\begin{document}

\maketitle

\begin{abstract}
The gravitational clustering of collisionless particles in 
an expanding universe is modelled using some simple physical ideas. I
show that it is possible to understand the nonlinear clustering
in terms of three well defined regimes: (1) linear regime; (2)
quasilinear regime which is dominated by scale-invariant radial infall
and (3) nonlinear regime dominated by nonradial motions and
mergers. Modelling each of these regimes separately I show how the
nonlinear two point correlation function can be related to the linear
correlation function in hierarchical models. This analysis leads to
results which are in good agreement with numerical simulations thereby
providing an explanation for numerical results. Using this model
and some simple extensions, it is possible to understand the transfer
of power from large to small scales and the behaviour of higher
order correlation functions. The ideas presented
here will also serve as a powerful analytical tool to investigate
nonlinear clustering in different models.
\end{abstract}

\section{Introduction}

Consider a collection $N$ point particles, interacting with each other
by the Newtonian  gravity, in an expanding background characterized by
a scale factor $a(t)$. What can we say about the time evolution of
such a system?

This problem is of considerable interest for several reasons. To begin
with, the behaviour of large number of particles interacting via
Newtonian gravity  poses a formidable challenge to the usual methods
of statistical mechanics (T. Padmanabhan, 1990). 
 So, purely from an academic point of view, this seems
to be a challenging but solvable problem.

Secondly,  this problem might even have some practical interest. There
is considerable evidence that the universe is dominated by
collisionless non-baryonic dark matter particles. In that case, they
will play a key role in the formation of large scale structures. If
the length scales of interest are (i) small compared to  Hubble radius
but (ii) large compared to the  scales at which non-gravitational
processes are significant, then the system of dark matter particles
constitutes an example in which the question raised in the first
paragraph becomes relevant. In fact, most of the work in this subject
has been inspired by considerations of structure formation. 

A brute force method for solving this problem relies on numerical
simulations. In such an approach one starts with  large number of
particles distributed nearly uniformly and calculates the future
trajectories by a suitable numerical algorithm. This approach, however, does not
lead to genuine understanding unless we supplement it with
some analytic modelling.
In this talk, I shall outline how one can make analytic
progress in the problem of nonlinear gravitational clustering and thereby
reproduce the key features of numerical simulations (Padmanabhan, 1996 a,b).

\section{Modelling the gravitational clustering }

Consider a system of particles distributed homogeneously,
on the average, with a mean density $\bar \rho (t)$. This uniform
density will cause an expansion of the universe and the proper
distance ${\bf r}  = a (t) \bf x $ between particles will increase
with time. If the distribution was not strictly uniform, then the
perturbations in the density will act as local centres of
clustering. A region with overdensity will accrete matter around it
while an underdense region will repel matter in its surroundings. As a
result, perturbations in  density will tend to grow and, when the
density contrast is of order unity, these cluster centres will exert
significant influence on the evolution. Particles in a highly
overdense regions will evolve essentially under  their own
self-gravity and will tend to form gravitationally bound
systems.

When these density perturbations are small, it is possible to study
their evolution using linear theory. But once the density contrast
becomes comparable to unity, linear perturbation theory breaks down
and one must use  N-body simulations to study the growth of
perturbations. While these simulations are of  some value in making
concrete predictions for specific models, they do not provide clear
physical  insight into the process of non-linear gravitational
dynamics. To obtain such an insight into  this complex problem, it is
necessary to model the gravitational clustering of collisionless
particles using simple physical concepts. I shall develop one such
model in this section, which - in spite of extreme simplicity -
reproduces the simulation results for hierarchical models fairly
accurately. Further, this model also provides insight into the
clustering process and can be modified to take into account more
complicated situations. 

The paradigm for understanding the clustering is based on the well
known behaviour of a spherically symmetric overdense region in the
universe. In the behaviour of such a region, one can identify
three different regimes of interest: (1) In the early stages of the
evolution, when the density contrast is small, the evolution is
described by linear theory. (2) Each of the spherical shells with 
an initial radius $x_i$ can be parametersed by a mass contained inside
the shell, $M(x_i),$ and the energy, $E(x_i)$ for the particular
shell. Each shell will expand to a maximum radius $x_{max}\propto
M/|E|$ and then turn around and collapse. Such a spherical collapse
and resulting evolution allows a self similar description (Filmore \&
Goldreich, 1984; Bertshinger, 1985) in which each shell acts as though
it has an effective radius proportional to $x_{max}$. This will be the
quasilinear phase. (3) The spherical evolution will break down during
the later stages due to several reasons. First of all, non radial
motions will arise due to amplification of deviations from spherical
symmetry. Secondly, the existence of substructure will influence the
evolution in a non-spherically symmetric way. Finally, in the real
universe, there will be merging of such clusters [each of which could
have been centres of spherical overdense regions in the begining]
which will again destroy the spherical symmetry. This will be the
nonlinear phase. 

The description given above is  sufficiently
well known that one may suspect it can not lead to any insight into
the problem. In particular, structures observed in the real universe
are hardly spherical. I will show that it is, however, possible to
model the above process in a manner which allows direct generalisation
to the real universe. 

To do this we will begin by studying the evolution of system 
starting from a gaussian initial fluctuations with an initial power
spectrum, $P_{in}(k)$. The fourier transform of the power spectrum 
defines the correlation function $\xi(a,x)$ where $a\propto t^{2/3}$
is the expansion 
factor in a universe with $\Omega=1$. It is more convenient for our
purpose to work with the
average correlation function inside a sphere of radius $x$, 
defined by
\begin{equation}
\bar{\xi}(a,x)\equiv {3\over x^3}\int^{x}_{0}\xi(a,y)y^2 dy  
\end{equation}
In the linear regime we have $\bar{\xi}_L(a,x)\propto
a^2\bar\xi_{in}(a_i,x)$. In the quasilinear and nonlinear regimes, 
we would like to have prescription which relates the exact $\bar
\xi$ to the mean correlation function calculated from the linear
theory. One might have naively imagined that $\bar\xi(a,x)$ should
be related to $\bar\xi_{L}(a,x)$. But one can convince oneself that
the relationship is likely to be nonlocal by the following analysis:

Recall that, the conservation of pairs of particles, gives an exact
equation satisfied by the correlation function (Peebles, 1980):
\begin{equation}
{\partial\xi \over\partial t}+{1\over ax^2}{\partial\over\partial
x}[x^2(1+\xi)v]=0\label{qpaircon}
\end{equation}
where $v(t,x)$ denotes the mean relative velocity of pairs at
separation $x$ and 
epoch $t$. Using the mean correlation function $\bar\xi$ and a
dimensionless pair velocity $h(a,x) \equiv - (v/\dot{a}x)$, equation
(\ref{qpaircon}) can be written as
\begin{equation}
({\partial\over\partial \ln a}-h{\partial\over\partial \ln x})\,\,\,
(1+\bar{\xi})=3h(1+\bar{\xi})
 \end{equation}
This equation can be simplified by first introducing the variables
\begin{equation}
A=\ln a,\qquad X=\ln x ,\qquad D(X,A) = \ln (1+\bar{\xi})
\end{equation}
in terms of which we have (Nityananda and Padmanabhan, 1994)
\begin{equation}
{\partial D\over\partial A}-h(A,X){\partial D\over\partial
X}= 3h(A,X)\label{qkey}
\end{equation}
Introducing further a variable
$F=D+3X$, (\ref{qkey}) can be written  in a remarkably simple form as
\begin{equation}
{\partial F\over \partial A}-h(A,X){\partial F\over\partial
X}=0\label{qfunrel}
 \end{equation}
The charecteristic curves to this equation - on which $F$ is a
constant - are determined by 
$(dX/dA)=-h(X,A)$ which can be integrated if $h$ is known. But note that
the charecteristics satisfy the condition 
\begin{equation}
F=3X+ D=\ln [x^3(1+\bar{\xi})]={\rm constant}
\end{equation}
or, equivalently,
\begin{equation}
x^3(1+\bar{\xi})=l^3\label{qxandl}
\end{equation}
where $l$ is another length scale. When the evolution is linear at all
the relevant scales, $\bar{\xi}\ll 1$ and $l\approx x$. As clustering
develops,
$\bar{\xi}$ increases and $x$ becomes considerable smaller than $l$.
It is clear that the behaviour of clustering at some scale $x$ is
determined by
the original {\it linear} power spectrum at the scale $l$ through the
``flow of information'' along the charesteristics.
This suggests that {\it we should actually 
try to express the true
correlation function $\bar\xi(a,x)$ in terms of the linear correlation
function $\bar\xi_L(a,l)$ evaluated at a different point}.

Let us see how we can do this starting from  the quasilinear regime.
Consider a region surrounding a density peak in the linear stage,
around which we expect the clustering to take place. It is well known
that density profile around this peak
can be described by
\begin{equation}
\rho(x)\approx\rho_{bg}[1+\xi(x)] 
\end{equation}
Hence the initial mean density contrast scales with the initial shell
radius $l$ as $\bar\delta_i 
(l)\propto\bar\xi_L(l)$ in the initial epoch, when linear theory
is valid. This shell will expand to a maximum radius of $x_{max}
\propto l/\bar\delta_i\propto l/\bar\xi_L(l)$. In  scale-invariant,
radial collapse, models 
each shell may be approximated as contributing with a effective radius
which is propotional to $x_{max}$. Taking
the final effective radius $x$ as proportional to $x_{max}$, the final
mean correlation function will 
be
\begin{equation}
\bar\xi_{QL}(x)\propto \rho\propto {M\over x^3}
\propto {l^3\over (l^3/\bar\xi_L(l))^3}\propto
\bar\xi_L(l)^3 
\end{equation}
That is, the final correlation function $\bar\xi_{QL}$ at $x$ is the cube of
initial correlation function at $l$ where $l^3\propto x^3
\bar\xi_L^3\propto x^3\bar\xi_{QL}(x).$ This is in the form demanded
by (\ref{qxandl}) if $\bar\xi\gg 1$. {\it Note that we did not assume that
the initial power
spectrum is a power law to get this result.} 

In case the initial power spectrum {\it is} a power law, with
$\bar\xi_{L}\propto x^{-(n+3)}$, then we immediately find that
\begin{equation}
\bar\xi_{QL}\propto x^{-3(n+3)/(n+4)}\label{qlndep}
\end{equation}
[If the  
correlation function in linear theory has the powerlaw form $\bar\xi_{L}
\propto x^{-\alpha}$ then the process desribed above changes the index
from $\alpha$ to $3\alpha/(1+\alpha)$. We shall comment more 
about this aspect later]. For the power law case, the
same result can be obtained by more explicit means. For
example, in power law models the energy of spherical shell with mean density $\bar\delta(x_i) \propto x_i^{-b}$ will scale
with its radius as  
$E\propto G \delta M(x_i)/x_i \propto G \bar\delta x^2_i \propto x_i^{2-b}$. Since $M\propto x_i^3$, it follows that the
maximum radius reached by the shell scales as $x_{max}\propto
(M/E)\propto x_i^{1+b}$. Taking the effective radius as
$x=x_{eff}\propto x_i^{1+b}$,  the final density scales as
\begin{equation}
\rho\propto {M\over x^3}\propto {x_i^3\over x_i^{3(1+b)}}
\propto x_i^{-3b}\propto x^{-3b/(1+b)}\label{basres}
\end{equation}
In this quasilinear regime, $\bar\xi$ will scale like the density and we get
$\bar\xi_{QL}\propto x^{-3b/(1+b)}$. 
The index $b$ can be related to
$n$ by assuming the the evolution starts at a moment when linear
theory is valid. Since the gravitational potential energy [or the kinetic
energy] scales as $E\propto x_i^{-(n+1)}$ in the linear theory, it
follows that $b=n+3$. This 
 leads to  the correlation function in the quasilinear regime, given by (\ref{qlndep}) .

If $\Omega=1 $ and the initial spectrum is a power law, then there is
no intrinsic scale in the problem. 
It follows that the evolution has to be self similar and
$\bar\xi$ can only depend on the combination $q=xa^{-2/(n+3)}$. This allows to
determine the $a$ dependence of $\bar\xi_{QL}$ by substituting $q$
for $x$ in (\ref{qlndep}). We find
\begin{equation}
\bar\xi_{QL}(a,x)\propto a^{6/(n+4)}x^{-3(n+3)/(n+4)}\label{qlax}
\end{equation}
We know that, in the linear regime, $\bar\xi = \bar\xi_L \propto a^2$. Equation 
(\ref{qlax}) shows that, in the quasilinear regime, $\bar\xi = \bar\xi_{QL} \propto a^{6/(n+4)}$. Spectra with $n < -1 $ grow faster than $a^2$, spectra with $n > -1 $ grow slower than $a^2$ and $n = -1 $ spectrum grows as $a^2$.

 Direct algebra shows that
\begin{equation}
\bar\xi_{QL}(a,x)\propto [\bar\xi_{L}(a,l)]^3\label{qlscal}
\end{equation}
reconfirming the local dependence in $a$ and nonlocal dependence
in spatial coordinate.
This result has no trace of original assumptions [spherical evolution,
scale-invariant spectrum ....] left in it and hence once 
would strongly suspect that it will have far general validity.

Let us now proceed to the third and nonlinear regime. If we ignore the
effect of mergers, then it seems reasonable that virialised systems 
should maintain their densities and sizes in proper coordinates, i.e.
the clustering should be ``stable". This
would require the correlation function to have the form $\bar\xi_{NL}
(a,x)=a^3F(ax)$. [The factor $a^3$ arising from the decrease in
background density].
From our previous analysis we expect this to be a function of
$\bar\xi_L(a,l)$ where $l^3\approx x^3\bar\xi_{NL}(a,x)$. Let us write
this relation as
\begin{equation}
\bar\xi_{NL}(a,x)=a^3F(ax)=U[\bar\xi_L(a,l)]\label{qtr} 
\end{equation}
where $U[z]$ is an unknown function of its argument which needs
to be determined. Since linear correlation function evolves as
$a^2$ we know that we can write $\bar\xi_L(a,l)=a^2Q[l^3]$
where $Q$ is some known function of its argument. [We are using
$l^3$ rather than $l$ in defining this function just for future
convenience of notation]. In our case $l^3=x^3\bar\xi_{NL}(a,x)
=(ax)^3F(ax)=r^3F(r)$ where we have changed variables from 
$(a,x)$ to $(a,r)$ with $r=ax$. Equation (\ref{qtr}) now reads
\begin{equation}
a^3F(r)=U[\bar\xi_L(a,l)]=U[a^2Q[l^3]]=U[a^2Q[r^3F(r)]]
\end{equation}
Consider this relation as a function of $a$ at constant $r$. Clearly
we need to satisfy $U[c_1 a^2]=c_2a^3$ where $c_1$ 
and $c_2$ are constants. Hence we must have
\begin{equation}
U[z]\propto z^{3/2}.
\end{equation}
Thus in the extreme nonlinear end we should have 
\begin{equation}
\bar\xi_{NL}(a,x)\propto [\bar\xi_{L}(a,l)]^{3/2}\label{qnlscl} 
\end{equation}
[Another way deriving this result is to note that if $\bar\xi=
a^3F(ax)$, then $h=1$. Integrating (\ref{qkey}) with appropriate boundary
condition leads to (\ref{qnlscl}) .]
Once again we did not need to invoke the assumption that the
spectrum is a power law. If it is a power law, then we get,
\begin{equation}
\bar{\xi}_{NL}(a,x)\propto a^{(3-\gamma)}x^{-\gamma};\qquad 
\gamma={3(n+3)\over (n+5)} 
\end{equation}

This result is based on the assumption of ``stable clustering" and
was originally derived by Peebles (Peebles, 1965). It can be directly
verified that the right hand side of this equation can be expressed in
terms of $q$ alone, as we would have expected. 

Putting all our results together, we find that the nonlinear mean
correlation function can be expressed in terms of the linear mean 
correlation function by the relation:
\begin{equation}
\bar \xi (a,x)=\cases{\bar \xi_L (a,l)&(for\ $\bar \xi_L<1, \, \bar
\xi<1$)\cr 
{\bar \xi_L(a,l)}^3 &(for\ $1<\bar \xi_L<5.85, \, 1<\bar \xi<200$)\cr
14.14 {\bar \xi_L(a,l)}^{3/2} &(for\ $5.85<\bar\xi_L, \, 200<\bar
\xi$)\cr}\label{hamilton} 
\end{equation}
The numerical coefficients have been determined by continuity
arguments. We have assumed the linear result to be valid upto
$\bar\xi=1$ and the 
virialisation to occur at $\bar\xi\approx 200$
which is result arising from the spherical model.  The exact values of
the numerical coefficients can be  obtained only from simulations.

The true test of such a model, of course, is N-body simulations and
remarkably enough, simulations are very well represented by relations
of the above form. The simulation data for CDM, for example, 
is well fitted by (Padmanabhan etal., 1996):
\begin{equation}
\bar \xi(a,x)=\cases{\bar \xi_L(a,l) &(for\ $\bar \xi_L<1.2, \, \bar
\xi<1.2$)\cr 
{\bar \xi_L(a,l)}^3 &(for\ $1<\bar \xi_L<5, \, 1<\bar \xi<125$)\cr
11.7 {\bar \xi_L(a,l)}^{3/2} &(for\ $5<\bar\xi_L,  125<\bar
\xi$)\cr}\label{qbagh} 
\end{equation}
which is fairly close to the theoretical prediction.
[The fact that numerical simulations show a correlation between
$\bar\xi(a,x)$ and $\bar\xi_L(a,l)$ was originally pointed out 
by Hamilton et al. (1991) who, however, tried to give a multiparameter
fit to the data. This fit has somewhat obscured 
the simple physical interpretation of the result though has the virtue
of being very accurate for numerical work.]

A comparison of (\ref{hamilton}) and (\ref{qbagh}) shows that the
physical processes 
which operate at different scales are well represented by our model.
In other words, the processes descibed in the quasilinear and nonlinear
regimes for an {\it individual} lump still models the {\it average}
behaviour of
the universe in a statistical sense. It must be emphasised that the key
point is the ``flow of information" from $l$ to $x$ which is an exact
result.  Only when the results of the specific model are recast in
terms of suitably chosen variables, we get a relation which is of general
validity. It would have been, for example, incorrect to use spherical
model to obtain relation between linear and nonlinear densities at
the same location or to model the function $h$. 

It may be noted that to obtain the result in the nonlinear regime,
we needed to invoke the assumption of stable clustering which has
not been deduced from any fundamental considerations. In case
mergers of structures are important, one would consider this
assumption to be suspect (see Padmanabhan et al., 1996). We can,
however, generalise the above 
argument in the following manner: If the virialised systems have
reached  stationarity in the statistical sense, the function $h$
- which is the ratio between two velocities - should reach some
constant value. In that case, one can integrate (\ref{qfunrel}) and
obatin the result $\bar\xi_{NL}=a^{3h}F(a^hx)$. A similar argument
will now show that
\begin{equation}
\bar\xi_{NL}(a,x)\propto [\bar\xi_{L}(a,l)]^{3h/2}\label{qnlscl2}
\end{equation}
in the general case. For the power law spectra, one would get
\begin{equation}
\bar{\xi}(a,x)\propto a^{(3-\gamma)h}x^{-\gamma};\qquad 
\gamma={3h(n+3)\over 2+h(n+3)}
\end{equation}
Simulations are not accurate enough to fix the value of $h$; in
particular, the asymptotic value of $h$ could depend on $n$
within the accuracy of the simulations. It may be possible to
determine this dependence by modelling mergers in some simplified form.

If $h = 1$ asymptotically, the correlation function in the extreme
nonlinear end depends on the linear index $n$. One may feel that
physics at highly nonlinear end should be independent of the linear
spectral index $n$. This will be the case if the asymptotic value of
$h$ satisfies the scaling 
\begin{equation}
h = {3c \over n+3} 
\end{equation}
in the nonlinear end with some constant $c$. Only high resolution
numerical simulations can test this conjecture that $h(n + 3 ) = {\rm
constant}$. 

It is possible to obtain similar relations between $\xi(a, x)$ and
$\xi_L (a, l) $ in two dimensions as well. In 2-D the scaling
relations turn out to be

\begin{equation}
\bar \xi (a,x)\propto \cases{\bar \xi_L (a,l)&({\rm Linear}) \cr
\bar\xi_L(a,l)^2 &({\rm Quasi-linear})\cr
\bar\xi_L(a,l) &({Nonlinear}) \cr}
\end{equation}
For power law spectrum the nonlinear correction function will
$\bar\xi_{NL} (a, x) = a^{2 - \gamma} x^{-\gamma} $ with $\gamma = 2
(n + 2) / (n + 4)$. 

If we generalize the concept of stable clustering to mean constancy of
$h$ in the nonlinear epoch, then the correlation function will behave
as $\bar\xi_{NL} (a, x) = a^{2h}F(a^hx)$. In this case, if the
spectrum is a power law then the nonlinear and linear indices are
related to 
\begin{equation}
\gamma = {2h (n + 2) \over 2 + h (n + 2)}
\end{equation}
All the features discussed in the case of 3 dimensions are present
here as well. For example, if the asymptotic value of $h$ scales with
$n$ such that $h (n + 2 )= {\rm constant}$ then the nonlinear index
will be independent of the linear index. (Numerically it would be lot
easier to test this result in 2-D rather than in 3-D; work is in
progress to test these results). 

We shall now consider some applications and further generalisations of
our model.

\section{Critical Index and power transfer}

Given a model for the evolution of the power spectra in the quasilinear
and nonlinear regimes, one could explore whether 
evolution of gravitational clustering
possesses any universal charecteristics. For example one could ask
whether a complicated initial power spectrum will be driven to any
particular form of power spectrum in the late stages of the evolution.

One suspects that such a possibility might arise because of the following reason: We saw in the last section that [in the
quasilinear regime] spectra with $n<-1$ grow faster
than $a^2$ while spectra with $n>-1$ grow slower than $a^2$. This feature
could drive the spectral index to $n=n_c\approx -1$ in the quasilinear
regime irrespective of the initial index. Similarly, the index in
the nonlinear regime could be driven to $n\approx -2$ during the late time evolution. So the spectral indices $-1$ and $-2$ are some kind
of ``fixed points" in the quasilinear and nonlinear regimes. Speculating along
these lines, we would expect the gravitational clustering to lead to
a  ``universal" profile which scales as $x^{-1}$ at the nonlinear end changing over to $x^{-2}$ in
the quasilinear regime.

This effect can be understood better by studying the ``effective" index
for the power spectra at different stages of the evolution. These are plotted in figure 1. The three panels of figure 1 illustrate features related to the
existence of fixed points in a clear manner. In the top panel we have
plotted index of growth $n_a\equiv(\partial \ln\bar\xi(a,x)/\partial
\ln a)_x$ as a function of $\bar\xi$ in the quasilinear regime
obtained from our scaling relations. Curves
correspond to an input spectrum with index $n=-2,-1,1$. The dashed
horizontal line at $n_a=2$ represents the linear growth rate. An index
above this dashed horizontal line will represent a rate of growth faster than
linear growth rate and the one below will represent a rate which is
slower than the linear rate. It is clear that -- in the quasilinear
regime -- the curve for $n=-1$ closely follows the linear growth while
$n=-2$ grows faster and $n=1$ grows slower; so the critical index is
$n_c\approx -1$. The curves are based on the fitting formula due to
Hamilton et al, 1991.

The second panel of figure 1 shows the effective index $n_a$ as a
function of the index $n$ of the original linear spectrum at different
levels of nonlinearity labelled by $\bar\xi=1,5,10,50,100$. We see
that in the quasilinear regime, $n_a>2$ for $n<-1$ and $n_a<2$ for
$n>-1$.

The lower panel of figure 1 shows the slope $n_x = -3 - (\partial\ln
{\bar\xi} /\partial \ln{x})_a $ of $\bar\xi$ for different power law
spectra. It is clear that $n_x$ crowds around $n_c\approx -1$ in the
quasilinear regime. If perturbations grow by gravitaional instability,
starting from an epoch at which $\bar\xi_{initial}\ll 1$ at all
scales, then it can be shown that $n_x$ at any epoch must satisfy the
inequality
\begin{equation} 
n_x\le (3/\bar\xi).\label{qq}
\end{equation}
This bounding curve is shown by a dotted line in the figure. This
powerful inequality shows that regions of strong nonlinearity [with
$\bar\xi\gg 1$] should have effective index which is close to or less
than zero.

\begin{figure}
\epsfxsize=3.3truein\epsfbox[37 402 558 744]{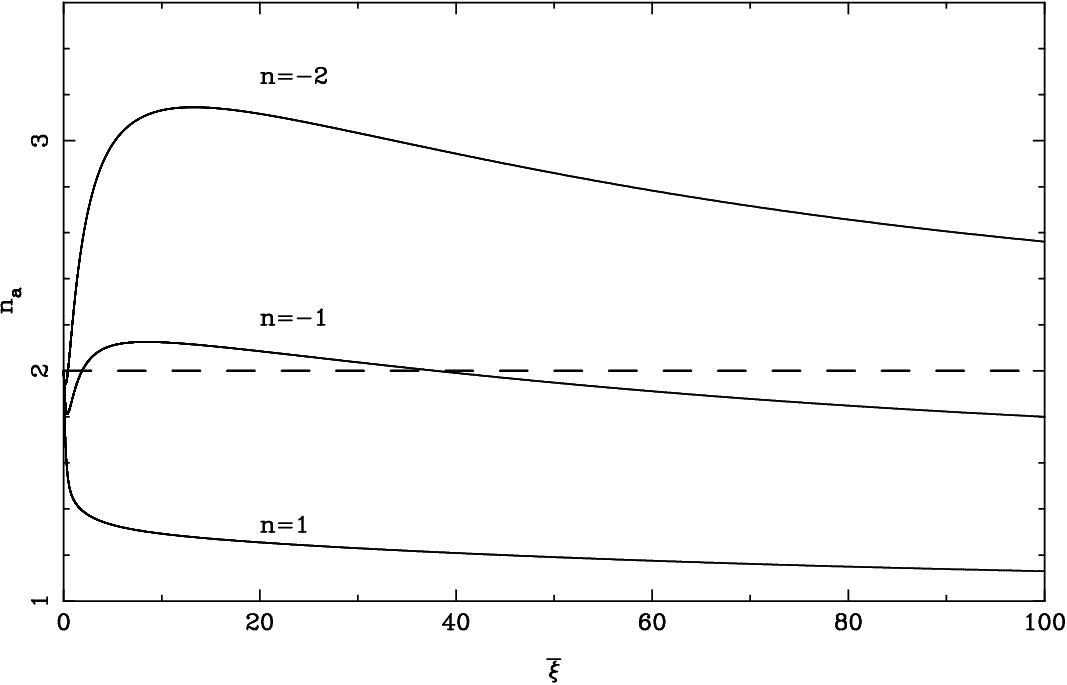}
\epsfxsize=3.3truein\epsfbox[37 399 552 737]{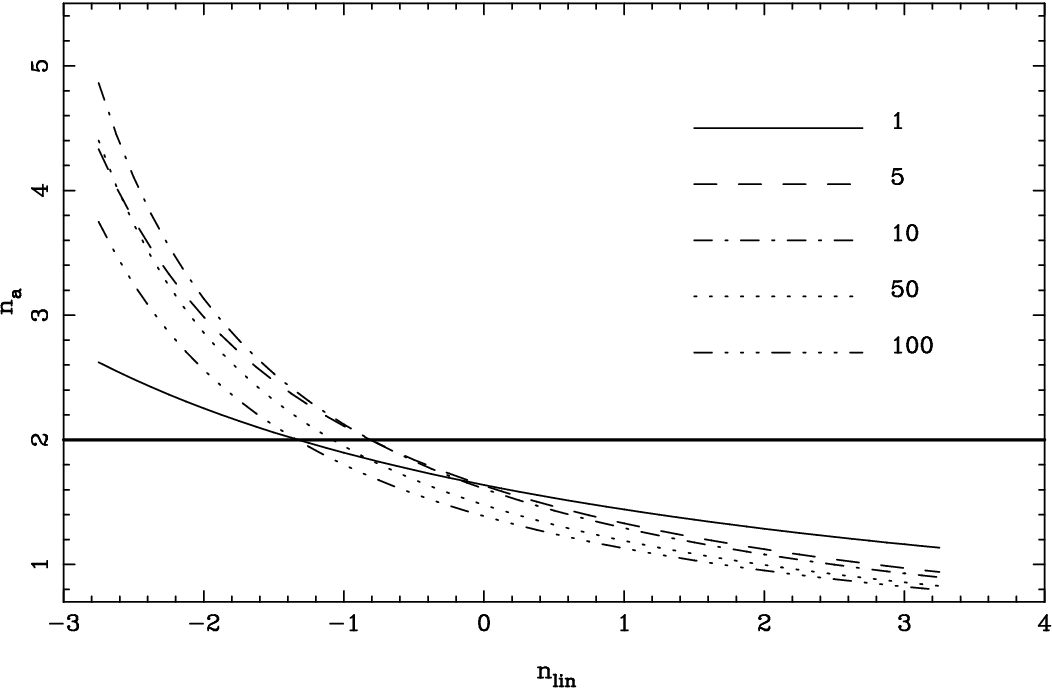}
\epsfxsize=3.3truein\epsfbox[37 399 552 737]{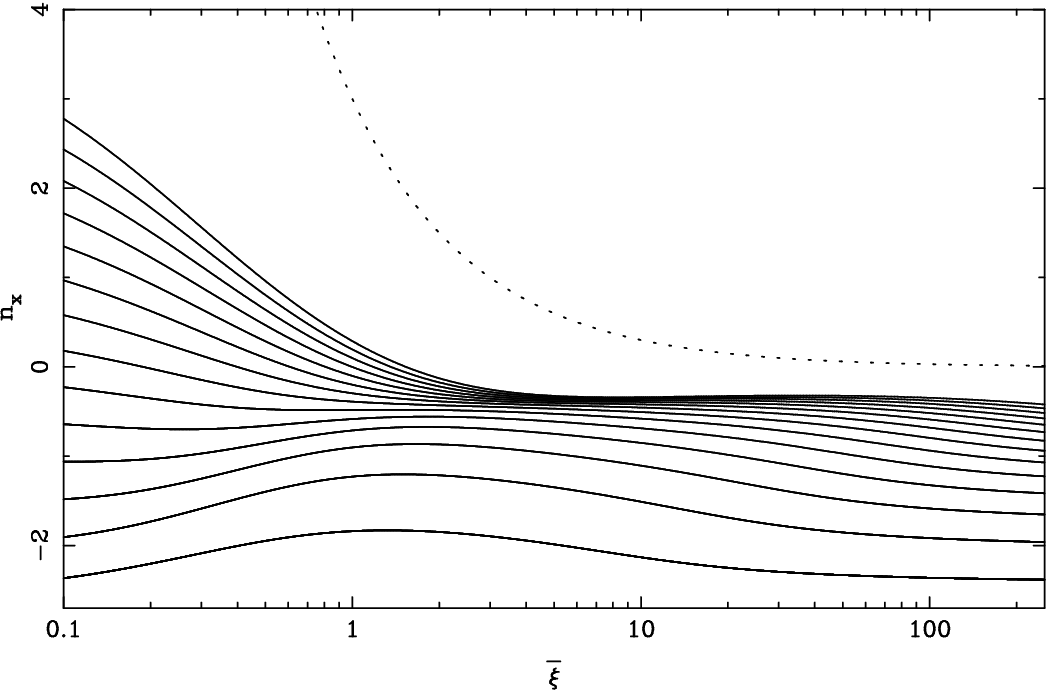}
\caption{The top panel shows exponent of rate of growth of density
fluctuations 
as a function of amplitude. We have plotted the rate of growth for
three scale invariant spectra $n=-2, -1, 1$. The dashed horizontal
line indicates the exponent for linear growth. For the range
$1<\delta<100$, the $n=-1$ spectrum grows as in linear theory; $n<-1$
grows faster and $n>-1$ grows slower. The second panel shows exponent
of rate of growth as a function of linear index of the power spectrum
for different values of $\bar\xi$ $( 1,5,10,50,100)$. These are
represented by thick, dashed, dot-dashed, dotted and the dot-dot-dashed 
lines respectively. It is clear that spectra with $n_{lin}<-1$
grow faster than the rate of growth in linear regime and $n_{lin}>-1$
grow slower. The lower panel shows the evolution of index
$n_x=-3-(\partial\ln {\bar\xi} /\partial \ln{x})_a$ with
$\bar\xi$. Indices vary from $n=-2.5$ to $n=4.0$ in steps of
$0.5$. The tendency for $n_x$ to crowd around $n_c=-1$ is apparent in
the quasilinear regime. The dashed curve is a bounding curve for the
index ($n_x < 3 /\bar\xi$) if perturbations grow via gravitational
instability.}
\end{figure}
The index $n_c=-1$ corresponds to the isothermal profile with
$\bar\xi(a,x)=a^2x^{-2}$ and has two interesting features to
recommend it as a candidate for fixed point:

(i) For $n=-1$ spectra each logarithmic scale contributes the same
amount of correlation potential energy.  If the regime is modelled by scale
invariant radial flows, then the kinetic energy will scale in the same
way. It is conceivable that flow of power leads to such an
equipartition state as a fixed point though it is difficult prove such
a result in any generality.

(ii) It can be shown that scale invariant spherical collapse will
change the density profile $x^{-b}$ with an index $b$ to another
profile with index $3b/(1+b)$. Such a mapping has a nontrivial fixed
point for $b=2$ corresponding to the isothermal profile and an index
of power specrum $n=-1$ (see Padmanabhan, 1996a).

These considerations also allow us to predict the nature of power
transfer in gravitational clustering. Suppose that, initially, the
power spectrum was sharply  
peaked at some scale
$k_0=2\pi/L_0$ and has a small width $\Delta k$. When the peak
amplitude of the spectrum is far less than unity, the evolution
will be described by linear theory and there will be no flow
of power to other scales. But once the peak approaches a value
close to unity, power will be generated at other scales due to nonlinear
couplings {\it even though the amplitude of perturbations in
these scales are less than unity}.
 Mathematically, this
can be understood from the evolution equations for the density contrast
- written  in fourier space - as :
\begin{equation}
\ddot\delta_{\bf k}+2{\dot a\over a}\dot\delta_{\bf k}=4\pi
G\bar\rho\delta_{\bf k} +Q \label{coupling}
\end{equation}
where $\delta_{\bf k}(t)$ is the fourier transform of the density
contrast, $\bar\rho$ is the background density and $Q$ is a nonlocal,
nonlinear function which couples the mode ${\bf k}$ to all other modes
${\bf k'}$ (Peebles, 1980). Coupling between different modes is
significant in two cases. The obvious case is one with $\delta_{\bf k}
\ge 1$. A more interesting possibility arises for modes with no
initial power [or exponentially small power]. In this case nonlinear
coupling provides the only driving terms, represented by $Q$ in
equation (\ref{coupling}). These generate power at the scale ${\bf k}$
through mode-coupling, provided power exists at some other scale. {\it
Note that the growth of power at the scale ${\bf k}$ will now be
governed purely by nonlinear effects even though $\delta_{\bf k} \ll
1$.} 

Physically, this arises along the following lines: If the initial
spectrum is sharply peaked at some scale $L_0$, first structures to
form  are voids with a typical diameter
$L_0$. Formation and fragmentation of sheets bounding the voids lead
to generation of power at scales $L<L_0$. First bound structures will then form
at the mass scale corresponding to $L_0$. In such a model, 
$\bar\xi_{\rm{lin}}$ at $L<L_0$  is nearly constant with an effective index of
$n\approx -3$. Assuming we can use equation (\ref{hamilton}) with the
local index in this case, we expect the power to grow very rapidly
as compared to the linear rate of $a^2$. [The rate of growth is $a^6$
for $n= -3$ and $a^4$ for $n=-2.5$.] Different rate of growth for
regions with different local index will lead to steepening of
the power spectrum and an eventual slowing down of the rate of
growth. In this process, which is the dominant one, 
 the power transfer is mostly
from large scales to small scales. [There is also a 
generation of the $k^4$ tail at large scales which we shall not discuss
here; see Bagla and Padmanabhan, 1996]. 

From our previous discussion, we would have expected such an evolution
to lead to a ``universal'' 
power spectrum with some critical index $n_c\approx -1$ 
for which the rate of growth is that of linear theory - viz.,
$a^2$. In fact, the same results should  hold even when there exists small
scale power; recent numerical simulations dramatically confirm this
prediction and show that - in the quasilinear
regime, with $1<\delta<100$ - power spectrum indeed has a universal slope [see Bagla and
Padmanabhan, 1996]. 

\section{Further generalizations}

The ideas presented here can be generalised in two obvious directions
(see Munshi and Padmanabhan, 1996): (i) By considering peaks of
different heights, drawn from an initial gaussian random field,
and averaging over the probability distribution one can obtain
a more precise scaling relation. (ii) By using a generalised ansatz for
higher order correlation functions, one can attempt to compute
the $S_N$ parameters in the quasilinear and nonlinear regimes. I
shall briefly comment on the results of these two generalisations.  

(i) The basic idea behind the model used in section 2  can be described
as follows: Consider the evolution of density perturbations starting from
an initial configuration, which is taken to be a realisation of a Gaussian
random field with variance $\sigma$. A region with initial density contrast $\delta_i$ will expand
to a maximum radius $x_f = x_i/ \delta_i$ and will contribute to the 
two-point correlation function an amount proportional to $(x_i/x_f)^3 = 
\delta_i^3$. The initial density contrast within a 
{\it randomly} placed
sphere of radius $x_i$ will be $ \nu \sigma (x_i)$ with a probability
proportional to $\exp (-\nu^2/2)$. On the other hand, the initial density 
contrast within a sphere of radius $x_i$, {\it centered around a peak in the 
density field} will be proportional to the two-point correlation function 
and will be  $\nu^2 \bar\xi (x_i)$ with a probability proportional to $\exp (-\nu^2/2)$. It follows that the contribution from a typical region will 
scale as  $ \bar \xi
\propto \bar \xi_i^{3/2}$ while  that from higher peaks will scale as $\bar \xi 
\propto \bar \xi_i^3$. In the quasilinear phase, most dominant contribution
arises from high peaks and we find the scaling to be  $\bar \xi_{QL} 
\propto \bar \xi_i^3$. The non-linear, virialized, regime is dominated by 
contribution from several typical initial regions and has the scaling
 $\bar \xi_{NL} 
\propto \bar \xi_i^{3/2}$. This was essentially the result obtained in section 2 except that we took  $\nu = 1$. 
To take into account the statistical fluctuations of the initial Gaussian
field we can average over different $\nu$ with a Gaussian probability 
distribution.

Such an analysis leads to the following result. The relationship between
$\bar \xi(a,x)$ and $\bar \xi_{L}(a,l)$ becomes 
\begin{equation}
\bar \xi (a,x) = A\left[ \bar \xi_{L} (a,l)\right]^{3h/2}; A = \left( {2\over \lambda} \right)^{3h \over 2} \left[ {\Gamma\left({\alpha + 1\over 2}\right) \over 2\sqrt{\pi}}\right]^{3h/ \alpha} \label{e1}
\end{equation}
where
\begin{equation}
\alpha = {6h\over 2+h(n+3)} \label{e2}
\end{equation}
and $\lambda \approx 0.5$ is the ratio between the final virialized radius and the radius at turn-around. In our model, $h=2$ in the quasi-linear regime,  and $h=1$ in the non-linear regime.  However, the above result holds for
any other value of $h$. Equation (\ref{e1}) shows that the scaling relations
 (\ref{hamilton}) acquire  coefficients which depend on the spectral index
$n$ when we average over peaks of different heights. This effect is seen 
in simulations and equation (\ref{e1}) correctly accounts for the numerical 
results (Munshi and Padmanabhan, 1996).  

(ii) In attempting to generalize our results to higher order correlation functions,
it is important to keep the following aspect in mind. The $N$th order correlation function will involve $N-1$ different length scales.
To make progress, one needs to assume that,  although there are 
different length scales present in reduced n-point correlation function,
all of them have to be roughly of the same order to give a significant contribution.  If the correlation functions are described
by a single scale, then a natural generalisation of equation (\ref{qxandl}), will be
\begin{equation}
\bar \xi_N \approx \langle x_i^{3(N-1)} \rangle
/ x^{3(N-1)}\end{equation}
Given such an ansatz for the $N$ point correlation function, one can compute
the $S_N$ coefficients defined by the relation $S_N \equiv \bar \xi_N / \bar \xi_2^{N-1}$ in a straightforward manner. We find that
\begin{equation}
S_N = \left( 4\pi \right)^{(N-2)/2} {\Gamma \left( {\alpha (N-1) +1 \over 2} \right) \over \left[ \Gamma \left({\alpha +1\over 2}\right) \right]^{N-1}}
\label{sn}
\end{equation}
where $\alpha$ is defined in equation (\ref{e2}). Given the function $h(\bar \xi)$, this equation allows one to compute (approximately) the value of 
$S_N$ parameters in the quasi-linear and non-linear regimes. In our model
$h =2$ in the quasi-linear regime and $h =1$ in the non-linear regime. The 
numerical values of $S_N$ computed for different power spectra agrees 
reasonably well with simulation results. (For more details, see Munshi and 
Padmanabhan, 1996.)

\end{document}